\documentclass[10pt,english]{article}
\usepackage[T1]{fontenc}
\usepackage[latin9]{inputenc}
\usepackage[letterpaper]{geometry}
\usepackage{color}
\usepackage{amsthm}
\usepackage{amsmath}
\usepackage{amssymb}

\makeatletter

\let\SF@@footnote\footnote
\def\footnote{\ifx\protect\@typeset@protect
    \expandafter\SF@@footnote
  \else
    \expandafter\SF@gobble@opt
  \fi
}
\expandafter\def\csname SF@gobble@opt \endcsname{\@ifnextchar[
  \SF@gobble@twobracket
  \@gobble
}
\edef\SF@gobble@opt{\noexpand\protect
  \expandafter\noexpand\csname SF@gobble@opt \endcsname}
\def\SF@gobble@twobracket[#1]#2{}

\newcommand{\lyxaddress}[1]{
\par {\raggedright #1
\vspace{1.4em}
\noindent\par}
}

\@ifundefined{date}{}{\date{}}

\usepackage{multirow}

\def\frontmatter@abstractheading{}

\usepackage{babel}

\makeatother

\usepackage{babel}
\begin{document}

\title{A Note on Quantum States and Observables in Psychological Measurements}

\author{Ehtibar N. Dzhafarov}

\maketitle

\lyxaddress{}

\lyxaddress{\begin{center}
\vspace{-7ex}
Purdue University, USA (ehtibar@purdue.edu) 
\par\end{center}}
\begin{abstract}
The problem considered is how to map the concepts of Quantum Theory
(QT) to elements of a psychological experiment. The QT concepts are
``measurement,'' ``state,'' and ``observable''. The elements
of a psychological experiment are trial, stimulus, instructions, questions,
and responses. 

\end{abstract}

\section{Introduction}

This note can be viewed as an extensive commentary to {[}KBDB{]}. 

Quantum Theory (QT) operates with observables and states. The problem
we consider here is how to map these concepts to those describing
a psychological experiment. 

On a very general level, QM accounts for the probability distributions
of measurement results using two kinds of entities, called \emph{observables}
$A$ and \emph{states} $\psi$ (of the system on which the measurements
are made). We assume that measurements are performed in a series of
consecutive trials numbered $1,2,\ldots$. In each trial $t$ the
experimenter decides what measurement to make (e.g., what question
to ask), and this amounts to choosing an observable $A$. The formulas
are 
\begin{equation}
\Pr\left[v\left(A\right)=v\textnormal{ in trial }t\,\vert\,\mbox{measurements in trials }1,\ldots,t-1\right]=F\left(\psi^{\left(t\right)},A,v\right).\label{eq: very general v}
\end{equation}
 
\begin{equation}
\psi^{\left(t+1\right)}=G\left(\psi^{\left(t\right)},A,v\right).\label{eq: very general rho}
\end{equation}

\textcolor{black}{
\begin{equation}
\psi_{\Delta}^{\left(t+1\right)}=H\left(\psi^{\left(t+1\right)},\Delta\right),\label{eq: very general evolution}
\end{equation}
}

In a psychological experiment we basic constituents of trials are
instructions/questions $q$ (specifying, among other things, the allowable
responses), and stimuli $s$.

\section{{[}KBDB{]} Approach}

In the approach adopted in {[}KBDB{]} stimuli and questions together,
($q,s$), determine observables. The state before the experiment is
generally undetermined, although influenced by the general instructions.
The state before every other trial is determined by (\ref{eq: very general rho})
and (\ref{eq: very general evolution}). The pair $\left(q,s\right)$
constitutes an \emph{input} to the system. 

For instance, a detection experiment in psychophysics is traditionally
considered as involving a single question ($q=$``does the stimulus
have the target property, yes or no?'') and two stimuli, the ``empty''
one $s=a$ and the target one $s=b$. We assume in {[}KBDB{]} that
the input $\left(q,s\right)$ in each trial uniquely determines the
observable $S$. Since only $s$ varies, we can denote the observables
by $A$ (corresponding to $a$) and $B$ (corresponding to $b$),
with two values each. But, of course, if the question were different
(e.g., $q=$``does the stimulus have the target property, yes or
no or uncertain?''), the observables would be different (although
there will still be two of them). The psychophysical analysis of such
an experiment consists in identifying the hit-rate and false-alarm-rate
functions (conditioned on the previous stimuli and responses) 
\begin{equation}
\begin{array}{c}
\Pr\left[v\left(A\right)=1\textnormal{ in trial }t\,\vert\,\mbox{measurements in trials }1,\ldots,t-1\right]=F\left(\psi^{\left(t\right)},A,1\right),\\
\Pr\left[v\left(B\right)=1\textnormal{ in trial }t\,\vert\,\mbox{measurements in trials }1,\ldots,t-1\right]=F\left(\psi^{\left(t\right)},B,1\right).
\end{array}
\end{equation}
The learning (or sequential-effect) aspect of such analysis consists
in identifying the function 
\[
\psi^{\left(t+1\right)}=G\left(\psi^{\left(t\right)},S,v\right),\; S\in\left\{ A,B\right\} ,v\in\left\{ 0,1\right\} ,
\]
combined with the ``pure'' inter-trial dynamics (\ref{eq: very general evolution}).

By contrast, an opinion-polling experiment is usually considered as
involving several questions $q$ and no sensory stimuli. Thus, in
one of Moore's polls we have $q=a=$``Is Bill Clinton honest, yes
or no?'', and $q=b=$``Is Al Gore honest, yes or no?''. The inputs
therefore are $a$ and $b$, and the corresponding observables are
$A,B$, with two values each. The analysis, formally, is precisely
the same as above:
\[
\begin{array}{c}
\Pr\left[v\left(A\right)=1\textnormal{ in trial }t\,\vert\,\mbox{measurements in trials }1,\ldots,t-1\right]=F\left(\psi^{\left(t\right)},A,1\right),\\
\Pr\left[v\left(B\right)=1\textnormal{ in trial }t\,\vert\,\mbox{measurements in trials }1,\ldots,t-1\right]=F\left(\psi^{\left(t\right)},B,1\right).
\end{array}
\]

\section{{[}BB{]} Approach}

In this approach one strictly distinguishes questions from stimuli
and assume that questions are mapped into observables while stimuli
are mapped into states. So, in the detection experiment (with a single
question $q$ and two stimuli $a,b$) there is a single observable
$Q$ and two states $\psi_{a}$ and $\psi_{b}$: 
\begin{equation}
\begin{array}{c}
\Pr\left[v\left(Q\right)=1\textnormal{ in trial }t\textnormal{ with }a\,\vert\,\mbox{measurements in trials }1,\ldots,t-1\right]=F\left(\psi_{A},Q,1\right),\\
\Pr\left[v\left(Q\right)=1\textnormal{ in trial }t\textnormal{ with }b\,\vert\,\mbox{measurements in trials }1,\ldots,t-1\right]=F\left(\psi_{B},Q,1\right).
\end{array}
\end{equation}
The dynamics of the states here, (\ref{eq: very general rho})-(\ref{eq: very general evolution}),
are irrelevant, because whatever the transformation $\psi'_{s}=G\left(\psi_{s},Q,1\right)$,
the next state will be reset by the next stimulus into $\psi_{a}$
or $\psi_{b}$.

In the opinion-polling experiment the two approaches coincide, because
the input there consists of a question only.

\section{Comparison}

Consider the situation when, in the detection paradigm, a stimulus
$a$ is repeated in trials 1 and 2. In the {[}KBDB{]} approach we
have
\begin{equation}
\Pr\left[v\left(A\right)=1\textnormal{ in trial }1\,\vert\right]=F\left(\psi^{\left(1\right)},A,1\right),
\end{equation}
where $\psi^{1}$ is the initial state in which the participant is
at the start of trial 1 (due to her ``preparation'' by the pre-experiment
experience and by general instructions). The state then is transformed
into
\begin{equation}
\psi^{\left(2\right)}=G\left(\psi^{\left(1\right)},A,1\right),
\end{equation}
and then into
\begin{equation}
\psi_{\Delta}^{\left(2\right)}=H\left(\psi^{\left(2\right)},\Delta\right)
\end{equation}
in the interval $\Delta$ between the two trials. The next (conditional)
probability of responding 1 is 
\begin{equation}
\Pr\left[v\left(A\right)=1\textnormal{ in trial }2\,\vert\, v\left(A\right)=1\textnormal{ in trial }1\right]=F\left(\psi_{\Delta}^{\left(2\right)},A,1\right).
\end{equation}
Clearly, the relationship between $F\left(\psi^{\left(1\right)},A,1\right)$
and $F\left(\psi_{\Delta}^{\left(2\right)},A,1\right)$ is complex.
In particular, they need not coincide.

In the {[}BB{]} approach, however, the situation is much simpler.
We have in the first trial
\begin{equation}
\Pr\left[v\left(Q\right)=1\textnormal{ in trial }1\textnormal{ with }a\,\right]=F\left(\psi_{A},Q,1\right),
\end{equation}
and then, irrespective of how $\psi_{A}^{\left(1\right)}$ transforms
as a result of this measurement and between the two trials, in the
next trial we have
\begin{equation}
\Pr\left[v\left(Q\right)=1\textnormal{ in trial }2\textnormal{ with }a\,\vert\, v\left(Q\right)=1\textnormal{ in trial }1\textnormal{ with }a\,\right]=F\left(\psi_{A},Q,1\right).
\end{equation}
We see that the two probabilities must coincide. This is definitely
not what happens empirically (see {[}ACK{]}). 

The {[}BB{]} approach therefore has to be modified. Thus, one might
assume that stimulus $s$ determines the state $\psi_{s}$ not uniquely,
but depending on the previous state too:
\[
\psi'_{s}=K\left(s,\psi\right).
\]
This could save the approach, but would introduce a mechanism other
than described by the QT generalizations (\ref{eq: very general v})-(\ref{eq: very general rho})-(\ref{eq: very general evolution}).

\section{Logical Problems With the {[}BB{]} Approach%
\footnote{The argument presented in this section was suggested to me by Harald
Atmanspacher in a conversation we had in April 2014.%
}}

Even if by means of some extraneous to QT considerations one could
make the {[}BB{]} approach work, it would still encounter the logical
difficulty: it is not clear how to distinguish stimuli from questions. 

Thus, in the opinion polling (say, in the Moore's Clinton-Gore one
{[}M{]}), suppose that the respondents are first instructed ``We
will show you a picture of a well known politician: tell us whether
you trust him/her, yes or no.'' This would amount to a single question
$Q$. Then the pictures of Clinton and of Gore would amount to two
stimuli $a,b$. Intuitively, the results of this procedural modification
need not dramatically change the outcomes. 

And in the {[}KBDB{]} approach it does not: the input is still essentially
the same, consisting of a question specifying allowable responses
and of the variable identifier of the question's target (the spoken
or written word ``Clinton'' is not much different from Clinton's
photograph).

In the {[}BB{]} approach, however, the procedural modification in
question would amount to change from dealing with two observables
and with states varying according to projection-evolution rules to
dealing with a single observable and with states forced by the photographs.

\section{Conclusion}

Our analysis shows that {[}BB{]} is more problematic than {[}KBDB{]}.
No doubt, {[}BB{]} can be modified in many ways, but it seems that
{[}KBDB{]} is a more straightforward and general application of QT,
unifying psychophysical, opinion polling, and quantum physical considerations.

\paragraph*{Acknowledgments}

This work was supported by NSF grant SES-1155956. I am grateful to
Harald Atmanspacher for discussing with me the issues related to this
note.

\end{document}